\documentclass[aps,pre,showpacs,twocolumn,english]{revtex4}
\usepackage{bm}
\usepackage{graphicx}
\bibstyle{approve.bib}
\usepackage{setspace}
\usepackage[]{fontenc}
\usepackage[latin1]{inputenc}

\makeatletter

\usepackage{amsthm,amssymb}
\theoremstyle{plain}

\usepackage{babel}

\makeatother

\begin{document}
\title{Mean-field vs. stochastic models for transcriptional regulation}
\author{R. Blossey}
\affiliation{Biological Nanosystems, 
Interdisciplinary Research Institute, Lille University of Science and Technology,  USR 3078 CNRS,
Parc Scientifique de la Haute Borne, 50, Avenue Halley, F-59658 Villeneuve d'Ascq, France}
\author{C. V. Giuraniuc}
\affiliation{Biological Nanosystems, 
Interdisciplinary Research Institute, Lille University of Science and Technology, USR 3078 CNRS, 
Parc Scientifique de la Haute Borne, 50, Avenue Halley, F-59658 Villeneuve d'Ascq, France}

\date{\today}

\begin{abstract}
We introduce a minimal model description for the dynamics of transcriptional regulatory networks.
It is studied within a mean-field approximation, i.e., by deterministic ode's representing 
the reaction kinetics, and by stochastic  simulations employing the Gillespie algorithm. 
We elucidate the different results both approaches can deliver, depending on the 
network under study, and in particular depending on the level of detail retained in the respective 
description. Two examples are addressed in detail: the repressilator, a  
transcriptional clock based on a three-gene network realized experimentally in {\it E. coli}, 
and a bistable two-gene circuit 
under external driving, a  transcriptional network motif recently proposed to play a role in cellular development.
\end{abstract}

\maketitle

\section{Introduction}

Mathematical models for the dynamics of transcriptional regulation are traditionally 
formulated either in terms of ordinary differential equations \cite{goldbeter96,fall00}, 
or by purely stochastic models, based on Master equations \cite{vankampen} or 
by using the Gillespie algorithm \cite{gillespie77}. Both the deterministic and stochastic 
descriptions average out spatial degrees of freedom and hence are more 
similar to each other than is often acknowledged. In recent years,  a discussion 
has started on the effect of stochasticity on gene regulatory processes; exemplary studies are
\cite{vilar02,swain02,elowitz02,paulsson04,paulsson05}. Indeed,
already the fact that molecules involved in regulatory processes often exist only in 
small copy numbers can be significant for the dynamics of a given regulatory circuit,
and stochastic effects like bursting may have an important role for cellular function 
\cite{paulsson05}.

Models of regulatory dynamics suffer also from another problem which is the
lack of precise knowledge of reaction rates. Building dynamic models for a
large number of network elements can induce further arbitrariness due to a lack of
detailed knowledge of the interaction mechanisms involved. Approaches that
aim to describe larger networks are often deliberately reductionist to become
computationally tractable (see, e.g., \cite{dejong03}, building on pioneering work
by Glass, Kauffman and Thomas \cite{glass73,thomas73}), 
and the result of such computations 
can then only be called ``qualitative''. The effect of these reduction schemes, 
which within a physics-based notion could also be subsumed under the notion of ``coarse-graining",
therefore often lacks clarity as to what effect the approximations/simplifications have, since a
general systematics is not available (an exemplary discussion of this issue can be found in 
\cite{bundschuh03}).

In this paper we address the question of what effect such a reduction scheme has on
the dynamics of a given regulatory network in a systematic way. For this we start from 
a minimal model  description for transcriptional regulatory networks which coarse grains as many 
regulatory layers as possible (although they could of course be added back in later). 
We note that this modeling philosopy is in contrast to the usual way models of transcriptional
regulation are built in which first all avaliable biochemical detail is considered and then
reduced by way of approximation (as, e.g., in \cite{mueller06,widder07} and many other
similar examples). 
We then formulate both a deterministic (mean-field) version and a stochastic version 
of the transcriptional dynamics. This approach allows us to study the dynamics of 
basically all fundamental classes of transcriptional networks relevant for prokaryotic 
organisms, although we only look at few-gene networks in detail here. 

The outline of the paper is as follows. We first develop the kinetic 
reactions involved in transcriptional regulation. Subsequently, we formulate the
corresponding deterministic and stochastic versions of the dynamics. A separate
section of the paper is devoted to the application of both schemes to commonly
encountered regulatory motifs \cite{alon06}. Two examples are presented in more 
detail since they display richer structure: the repressilator, a three-gene network of 
inhibiting gates which acts as a genetic clock, previously realized experimentally in {\it E. coli} \cite{elowitz00}, and a regulatory
motif with multiple inputs which was recently proposed to be relevant for regulatory
processes in development \cite{saka07}.  For all these systems, we compare the results of the
deterministic calculations and their stochastic counterparts and evaluate the role
different regulatory mechanisms play for the observed outcome.

\section{The gene gate model}

\subsection{The transcriptional reactions}

\begin{figure}
\includegraphics[width=0.45\textwidth]{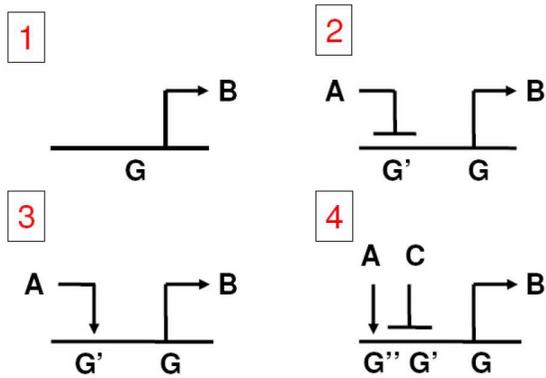}
\caption{(Color online) The four basic types of gene gates:
1) The null gate (a gate without control input); 
2) The neg gate (repression of transcription); 
3) The pos gate: activation of expression; 
4) the posneg gate: a multi-input gate with one activating and one repressing input.}
\end{figure}

Our minimal model for transcriptional regulation consists in the definition of a 
computational element for each regulatory element (i.e.,  transcribing gene), which we call a 
gene gate. The basic possible types of gene gates are sketched in Figure 1. 
Each gene gate is defined via its reaction kinetics. The `null gate' in Figure 1.1 is a gene in a state 
$G$ which produces a protein output $B$ at a rate $\varepsilon$, hence the kinetic reaction 
is written as
\begin{equation}
G \rightarrow_{\varepsilon} G + B\,.
\end{equation}
The protein output can be degraded according to the reaction
\begin{equation}
B \rightarrow_{\gamma} 0\, .
\end{equation}
In an abbreviating notation we call this gate element $ null(0;B)$ where inputs and
outputs are separated by the semicolon.

In the next step we add a regulatory input to the null gate. Figure 1.2 shows the resulting 
`neg gate' in which a transcription factor $A$ inhibits the production of protein $B$ upon 
binding. This is represented by the reaction
\begin{equation}
A + G \rightarrow_r G' + A\, .
\end{equation}
This reaction corresponds to the formation of a transcription factor-DNA complex with zero 
lifetime; such an intermediate with a finite lifetime can of course be introduced as well but 
is not necessary for a minimal model of gene networks. 

After this interaction, the gene in state $G'$ is blocked in transcription/translation. In order to allow transcription again the gate has to relax from its blocked state to its original transcribing state at a rate
$\eta$,
\begin{equation}
G' \rightarrow_{\eta} G 
\end{equation}
 to the state $G$ in which transcription at a basal rate $\varepsilon$ can occur. We call this gate the
$neg(A;B)$-gate. The relaxation process from $G'$ to $G$ models the fact that a gene generally is not transcribed immediately after the break-up of a transcription factor-DNA complex; also note that
within our minimal model of the gene gate, transcription and translation are lumped together.

Likewise we can model the activation of a gene upon binding of a transcription factor; 
Figure 1.3 shows the `pos gate'. The binding reaction is identical, but the gene in state $G'$
now behaves according to
\begin{equation}
G' \rightarrow_{\eta} G + B 
\end{equation}
where the rate $\eta > \varepsilon$, i.e. the transcription/translation rate upon activation is 
larger than the basal rate. This is the $pos(A;B)$-gate. 

Finally, Figure 1.4 shows a gate with multiple regulations which
is in fact a commonly encountered situation, see, e.g., the {\it E. coli} network
of transcriptional interactions reconstructed in \cite{babu03}.  
For the $posneg(A,C;B)$-gate we have to consider
three gene states, $G$, $G'$, and $G''$ with the reactions
\begin{equation}
C + G \rightarrow_{r_1} G' + C
\end{equation}
\begin{equation}
A + G \rightarrow_{r_2} G'' + A
\end{equation}
and the correponding relaxation reactions
\begin{equation}
G' \rightarrow_{\eta_1} G 
\end{equation}
\begin{equation}
G''\rightarrow_{\eta_2} G + B\, .  
\end{equation}
It is clear from this scheme that for each additional regulatory function, a binding
transcription factor and a corresponding gene state have to be introduced. 

Our minimal model obviously leaves out a number of regulatory levels such as
\begin{itemize}
\item complexation of transcription factors;
\item formation of the DNA-transcription factor complex;
\item DNA transcription and  RNA translation are lumped together.
\end{itemize}
These regulatory mechanisms can, of course, be added to the list of reactions given above,
and we will come back to this issue in the course of this paper.

\subsection{The mean-field equations} 

Having listed the transcriptional reactions we now define a continuum description based
on ordinary differential equations for the concentration of genes and proteins. 
We assume that the cell population can be considered 
as a `soup' containing the proteins as well as $N$ copies of the gene $G$. We
denote normalized concentrations by small letters $g \equiv [G]/N$, $ b \equiv [B]/N $ 
with $[G] \equiv \#G/V $ (likewise for $ [B]$) and keep the previous symbols for the kinetic constants 
(i.e., we include dependencies on cell volume $V$ and gene copy number $N$ where necessary; 
the difference to the kinetic reactions should be evident from the context). 
The two reactions of the null gate are then summarized by the ode
\begin{equation} \label{eqb}
\dot{b} = \varepsilon g - \gamma b\, .
\end{equation}
For the regulated genes, an equation for $g$ has to be added.
Since the $N$ gene gates present in our cell model have to be either in state $G$ or $G'$, 
one has the conservation law $ [G] +[ G'] = N$. From the normalization we have $ g + g' = 1$, 
and hence the {\it neg}-gate is described by the two odes, eq.(\ref{eqb}), and
\begin{equation} \label{eqg}
\dot{g} =  \eta g' - rga = \eta(1 - (1 + \nu a) g)\, ,
\end{equation}
where the conservation condition has been used, and $\nu \equiv r/\eta $.

The {\it pos}-gate (Figure 1.3) is governed by the ode's eq.(\ref{eqg}) and
\begin{equation}
\dot{b} = \varepsilon g + \eta g' - \gamma b = \eta  - (\eta - \varepsilon) g - \gamma b\, .
\end{equation}
Finally, we consider the case of multiple regulations of a single gene,
the simplest multi-input gate, the {\it posneg}-gate of  Figure 1.4 
with the three gene states, $G$, $G'$ and $G''$,
modifying the conservation condition to $g + g' + g'' = 1 $. We can
build up the gate reaction kinetics as before and obtain the system
of ode's
\begin{equation}
\dot{b} = \varepsilon b + \eta_2 g'' - \gamma b\, ,
\end{equation}
and
\begin{equation}
\dot{g}' =  - \eta_1 g' + r_1 g c\, ,
\end{equation}
\begin{equation}
\dot{g}'' =  - \eta_2 g'' + r_2 g a\, ,
\end{equation}
hence one has for $g$  the equation $ \dot{g} =  - (\dot{g}' + \dot{g}'') $ which follows 
from the conservation of gene states.

At this point we stress that we have only considered the case of binding of a single
protein $A$. In general, the binding of proteins is rather by multi-protein complexes
(dimers or higher), which is one way to give rise to a Hill coefficient $h$ when the
complexation reaction is considered an equilibrium (``fast") reaction \cite{weiss97}.
We could take this into account in our model by adding a corresponding 
complexation reaction in the reaction scheme. To be practical we here directly modify the ode 
equation of the gene by replacing $ a $ by $ a^h $ with $ h > 1 $ to cover this more general 
case; in what follows, we consider $h$ as a continously variable parameter. It is well-known 
that a Hill exponent $ > 1$ is essential for the dynamic behaviour of simple gene circuits 
\cite{cherry00}. 

For the stochastic simulations we employ the Gillespie algorithm which is equivalent to
the Chemical Master equation \cite{gillespie77}. We combine the Gillespie method with the 
stochastic $\pi$-calculus,  a process algebra originating in theoretical computer science 
\cite{milner99,pi,blossey06,blossey07,phillips07,phillips08}. For a brief introduction into
the main ideas of the calculus, see Appendix A.

\section{Examples}

\subsection{Basic circuits}
\begin{figure}[b]
\vspace*{-4cm}
\includegraphics[width=0.5\textwidth]{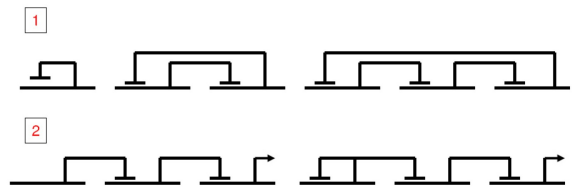}
\vspace*{-4cm}
\caption{(Color online) The two main classes of simple circuits: circular (1) and linear (2). Shown are only the repressive circuits; activatory circuits and mixtures of both types can be built in a similar fashion. 
Circuits shown in (1):
the autoinhibitive circuit, a bistable switch, the repressilator. Circuits in (2): a linear array and
a linear array with a head feedback: hence a mixture of a circular and a linear circuit.}
\end{figure}
We first discuss the elementary gene circuits that can be built from the above constructs. 
All simple transcriptional networks are either circular, linear or mixed circuits, see Figure 2. 
The archetypal loops are the autoinhibitory and autoactivatory loops.
The autoinhibitory loop $neg(a;a) $ is shown in Fig 2.1. The ode's governing its dynamics are
\begin{equation}
\dot{a} = \varepsilon g - \gamma a\, ,
\end{equation}
and
\begin{equation}
\dot{g} =  \eta g' - rga = \eta(1 - (1 + \nu a^h) g)\, .
\end{equation}
The natural first task is to look at nullclines and fixed-points. The nullcline of $g$ is determined 
by
\begin{equation}
g = \frac{1}{1 + \nu a^h}\, .
\end{equation}
If we have $\dot{g}/\eta \approx 0 $ and $\nu$ finite we can keep the
circuit near the nullcline of $g$. Inserting the nullcline condition into the equation for $a$ 
we find
\begin{equation}
\dot{a} = \frac{\varepsilon}{1 + \nu a^h} 
- \gamma a\, ,
\end{equation}
which is the common form of the Hill-type equation used in nonlinear dynamics
descriptions of gene networks. This turns out to be a general 
feature of the gene gate approach: near the nullclines of the gene gate
states, $ \dot{g} \approx  \dot{g'} \approx .... \approx 0 $, the circuit dynamics 
reduces to that of the standard Hill equations.
This feature has an immediate consequence for the fixed points.
The nuclline of $a $ is given by 
\begin{equation}
\frac{\varepsilon}{1 + \nu a^h} = \gamma a\, ,
\end{equation}
where the result for $g$ has been used, and we thus find the standard fixed-point
condition of the Hill equation for $a$. Since the left-hand side is a hyperbolic 
function in $a$, and the right-hand side is a linear function there is a unique
fixed-point of the circuit. 

The argument can be repeated for the autoactivatory loop $ pos(a;a) $ with the result
\begin{equation}
\dot{a} =  \eta  - \frac{\eta - \varepsilon}{1 + \nu a^h} - \gamma a
=  \frac{\varepsilon + r a^h}{1 + \nu a^h} - \gamma a \, ,
\end{equation}
which is the typical sigmoidal form of the activatory circuit. Therefore, we again find
that the fixed-points are given by a conditions akin to the standard Hill-type equations,
which for $ h > 1$ gives rise to three fixed-points.

The stability of the fixed-points in the gene networks 
is not affected by the presence of the genes. We illustrate
this for the bistable circuit composed of two neg-gtaes, 
$ neg(a;b)|neg(b;a) $, where the symbol $|$ denotes the composition of 
two gates, see Figure 2.1. The equations of th circuit read as
\begin{equation}
\dot{a} = \varepsilon g_a - \gamma a 
\end{equation}
and
\begin{equation}
\dot{g}_a = \eta (1 - (1 + \nu b^h)g_a)  
\end{equation}
and likewise for $a \leftrightarrow b$. 
As is well known \cite{cherry00}, the nonlinearity due to the Hill coefficient
is needed for the system in order to display the fixed-point structure of the bistable
switch; for a value of $h = 1$ as in our basic version of the gene gate model this is not the case. 
The stability of the fixed-points follows from the eigenvalues
of the matrix
\begin{equation}
\Gamma_{fp}
= 
\left(
\begin{array}{cccc}
-\gamma   & \varepsilon       & 0 	       & 0       \\
0                 & -\chi                   &  - \xi            & 0       \\
0		& 0			   & - \gamma  & 0       \\
- \xi            &  0                        & 0                 &- \chi  
\end{array}
\right)
\end{equation}
with
\begin{equation}
\chi \equiv \eta (1 + \nu a^h_i)\,\,,\,\,\,\, \xi \equiv \frac{r h a^{h-1}}{1 + \nu a^h}\,,
\end{equation}
Note that we are looking here at the stability of the {\it symmetric} fixed-point 
for which $\chi_1 = \chi_2 $, $\xi_1 = \xi_2 $. For the bistable switch, this is the
unstable fixed-point intervening between the two stable fixed-points,  and its
eigenvalues follow from the characteristic polynomial to $\Gamma_{fp}$,
\begin{equation}
(\gamma + \lambda)^2(\lambda + \chi)^2 = (\varepsilon \xi)^2\, .
\end{equation}
Taking the root of this equation, one finds four real eigenvalues, two of which
are negative, and two positive. The picture that emerges therefore is the
usual instability in the space of protein concentrations $a_1, a_2$, while the
genes do not contribute. 

We close this subsection by commenting on results from the stochastic simulations.
The basic loop- and linear circuits (negative, positive) show fixed-point  behaviour 
similar to their deterministic counterparts \cite{blossey06}.  For the bistable switch there is a
notable difference: 
as was recently shown based on a Master equation approach the stochastic dynamics 
of the bistable switch without cooperativity ($ h = 1 $) displays both bistability and 
switching \cite{biham06}. This behaviour is easily reproduced with our Gillespie approach, 
see Figure 3.
\begin{figure}[h]
\includegraphics[width=0.5\textwidth]{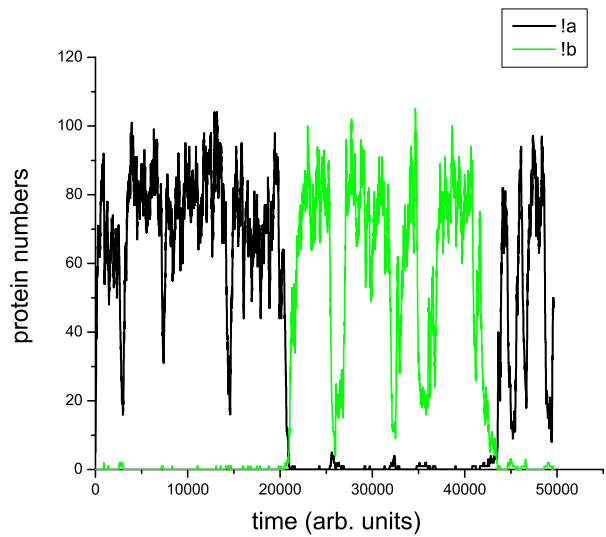}
\caption{(Color online) Switching in the stochastic bistable circuit without cooperativity. Simulation parameters
are: $r$ =1,  $\varepsilon$ = 0.4, $\eta$= 0.2,  $\gamma$ = 5 $\cdot 10^{-3}$. The insert
indicates output on the $\pi$-calculus channels $a!$, $b!$, equivalent to protein numbers.}
\end{figure}

Before moving on to richer examples, we draw a brief intermediate conclusion for the gene
gate model:
\begin{itemize}
\item if the deterministic gene circuit has a unique stable fixed-point, the genes are `irrelevant' variables in the sense that they do not alter the location of the fixed point. They do, however, affect the transient dynamics (see below);
\item the deterministic dynamics requires Hill-type nonlinearity in order to show bistability
and switching; for the stochastic dynamics, cooperativity is not needed.   
\end{itemize}

\subsection{The repressilator}

Clearly, the dynamics of the genes does affect the systems transients, and as such the genes can indeed have a profound influence on the dynamics,
as we now show. For this we look at a gene circuit whose stationary behaviour is not governed
by a simple fixed-point, but by a limit cycle: the repressilator. The repressilator is the three-gene negative-feedback loop shown in Figure 2.1;  
this system has been realized experimentally as a synthetic gene circuit in {\it E. coli} \cite{elowitz00}, and it has recently been the topic of various 
modeling papers, employing both deterministic and stochastic approaches, e.g.,  \cite{blossey06,mueller06,blossey07,biham07}. 

The nonlinear dynamics of the repressilator in the nullcline space of the gates is described by the ode
\begin{equation} \label{rep1} 
\dot{a} = \frac{\varepsilon}{1 + \nu b^h} - \gamma a 
\end{equation}
with the equations for $b$ and $c$ to be obtained from the 
permutations $(a \rightarrow b, b\rightarrow c)$ and  $(a \rightarrow c,  b\rightarrow a)$.
\begin{figure}
\includegraphics[width=0.45\textwidth]{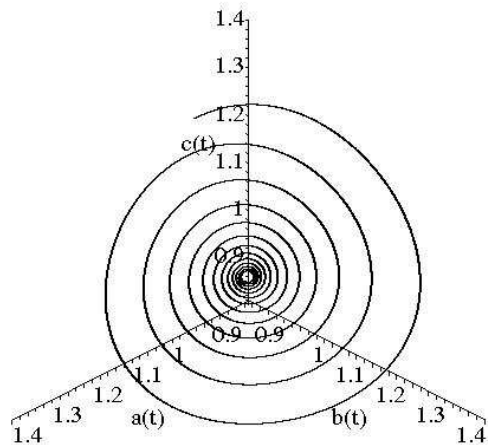}
\includegraphics[width=0.45\textwidth]{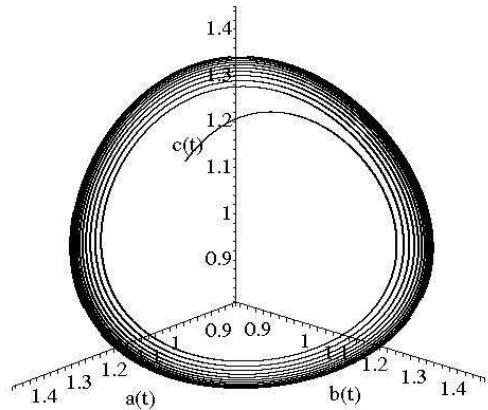}
\caption{Top: The repressilator dynamics without gene gates (fixed at the nullclines of the
gates) for the parameters  $r=1$,
$\gamma=0.1$, $\varepsilon = 0.3$, $\eta = 0.9$, $h=3$: the limit cycle is absent, the fixed-point
is stable. Bottom: Plot of the repressilator dynamics for the full system with identical parameters:  
the limit cycle persists in a wider range of parameters.}
\end{figure}
\begin{figure}
\includegraphics[width=0.55\textwidth]{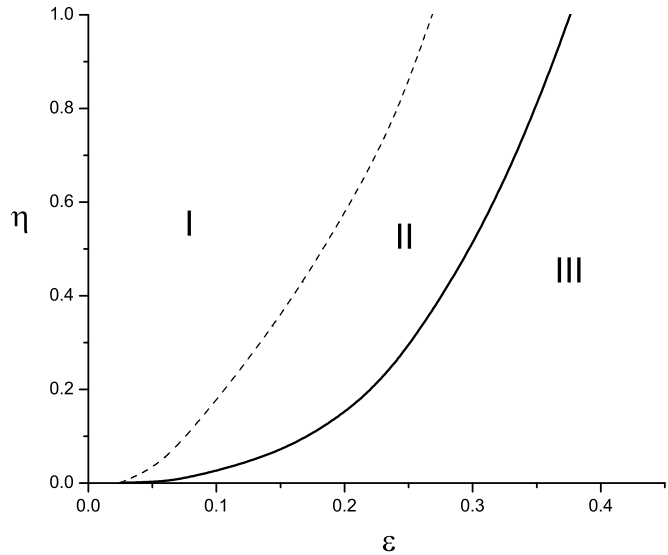}
\caption{Parameter regimes for the repressilator dynamics. 
I: stable fixed-point; II: stable fixed-point for the reduced system, limit cycle for the full system; III: limit cycle. Parameters are as in Figure 4.}
\end{figure}
Since all parameters are assumed equal the system has a symmetric fixed-point, 
$a=b=c\equiv \bar{a} $. Testing the stability of this fixed-point the stability matrix reads as
\begin{equation}
\Gamma^{fp}
= 
\left(
\begin{array}{ccc}
-\gamma   & - \kappa               & 0                     \\
0                 & -\gamma             &  - \kappa         \\
- \kappa     &  0                         & - \gamma  
\end{array}
\right)
\end{equation}
with $ \kappa = \varepsilon h \nu \bar{a}^{h-1}/(1 + \nu {\bar{a}}^h)^2 $.
The characteristic polynomial to this matrix is given by 
\begin{equation}
(\gamma + \lambda)^3 + \kappa^3 = 0\, ,
\end{equation}
so that the first eigenvalue is found to be
\begin{equation}
\lambda_1 = - (\gamma + \kappa)\,.
\end{equation}
The two others are given by
\begin{equation}
\lambda_{2,3} = - \gamma + \frac{\kappa}{2} \pm i \frac{\kappa}{2}\sqrt{3}\,.
\end{equation}
The condition for a Hopf-bifurcation therefore is
\begin{equation}
\frac{\kappa}{2} = \gamma\,.
\end{equation}
Making use of the fixed-point conditions one finds the relation
\begin{equation}
\bar{a} = \left(1 - \frac{2}{h}\right)\frac{\varepsilon}{\gamma} 
\end{equation}
and hence the condition on the Hill-exponent $ h > 2 $
for the circuit in order to have a stable limit cycle.

The stability analysis of this fixed-point can be carried out analytically for the full 
gene gate circuit, i.e. keeping both the transcription factors and the three genes as
dynamic variables. By symmetry, in fact, the calculation works for a circular circuit of $n$ genes.
The calculation amounts to generalize eq.(26) so that 
\begin{equation}
(\gamma + \lambda)^n(\lambda + \chi)^n + (-1)^{n - 1}(\varepsilon \xi)^n = 0\, .
\end{equation}
with $n=3$ for the repressilator. This fixed-point condition is
formally equivalent to that of the ``leaky'' repressilator discussed in
 \cite{mueller06}, for which a condition $ h > 4/3 $ was established. 
Within the full gene gate  dynamics, the condition on $h$ is thus weakened: the 
repressilator already oscillates for Hill exponent values less than two.  
Even for the case $ h = 3$, e.g., when both the full and the restricted system show 
oscillatory behaviour, the presence of the gene dynamics enlarges the oscillatory 
region in the space of protein concentrations. The stability of the limit cycle
in the space of parameters $(\varepsilon,\eta)$ is summarized in Figure 5.

By contrast, the stochastic repressilator without cooperativity displays a limit cycle
behaviour, as shown in Figure 6 (top).  
\begin{figure}
\begin{center}
\includegraphics[width=0.6\textwidth]{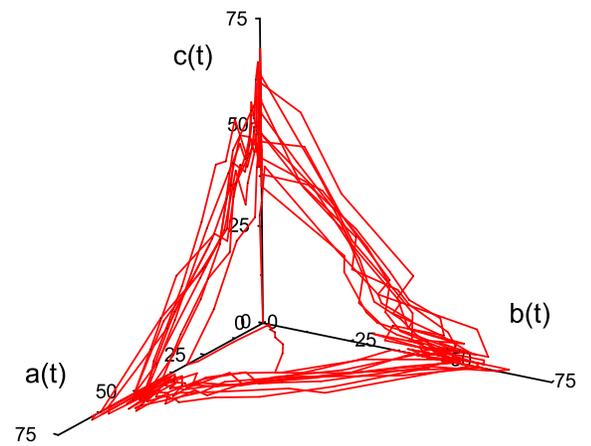}
\includegraphics[width=0.5\textwidth]{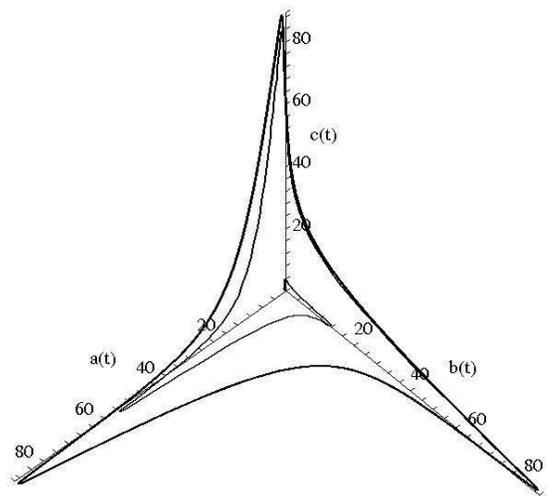}
\end{center}
\vspace{-2cm}
\caption{(Color online) Top: The limit cycle of the stochastic repressilator. Simulation parameters
are: $r = r_p $ = 1,  $\varepsilon$ = 0.1, $\eta$= $10^{-2}$,  $\gamma$ =  $10^{-3}$.
Bottom: the deterministic version for comparison (reduced system in region III of Figure 5, 
parameters identical to the stochastic version, with $ h = 3$).}
\end{figure}
The limit cycle appears as a symmetric triangle in the space of transcription factor concentrations
$(a,b,c)$. The triangle is somewhat `fuzzy', reflecting the fluctuating nature of the
concentrations. This fuzziness can be reduced by increasing the space of variables
in the system.
In a recent study, the effect of an inclusion of transcription factor cooperativity (dimerization
and higher), or an inclusion of explicit RNA transcription and protein translation was studied.
It was found that all these mechanisms regularize the oscillatory behaviour  \cite{blossey07}
and render the limit cycle less `fuzzy'. Analogous findings for circadian clocks were
reported earlier \cite{gonze02,gaspard02}.
The corresponding limit cycle for the deterministic dynamics
of the reduced system is shown in the bottom graph. Here again a Hill coefficient $h=3$ has been
assumed.

\section{Multi-input gates}

\subsection{A rewired repressilator}

The `stabilizing' effect due to the presence of the gene gates persists in the presence of 
multiple inputs, in fact, in can even be reinforced. We observed this when considering 
a rewired repressilator shown in Figure 7, in which an additional activatory loop has been 
added so that we have
\begin{equation}
neg(c;b)|posneg(c,b;a)|neg(a;c)
\end{equation}
\begin{figure}
\begin{center}
\includegraphics[width=0.45\textwidth]{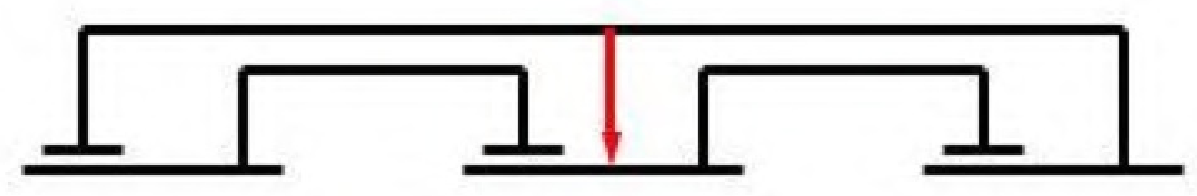}
\includegraphics[width=0.52\textwidth]{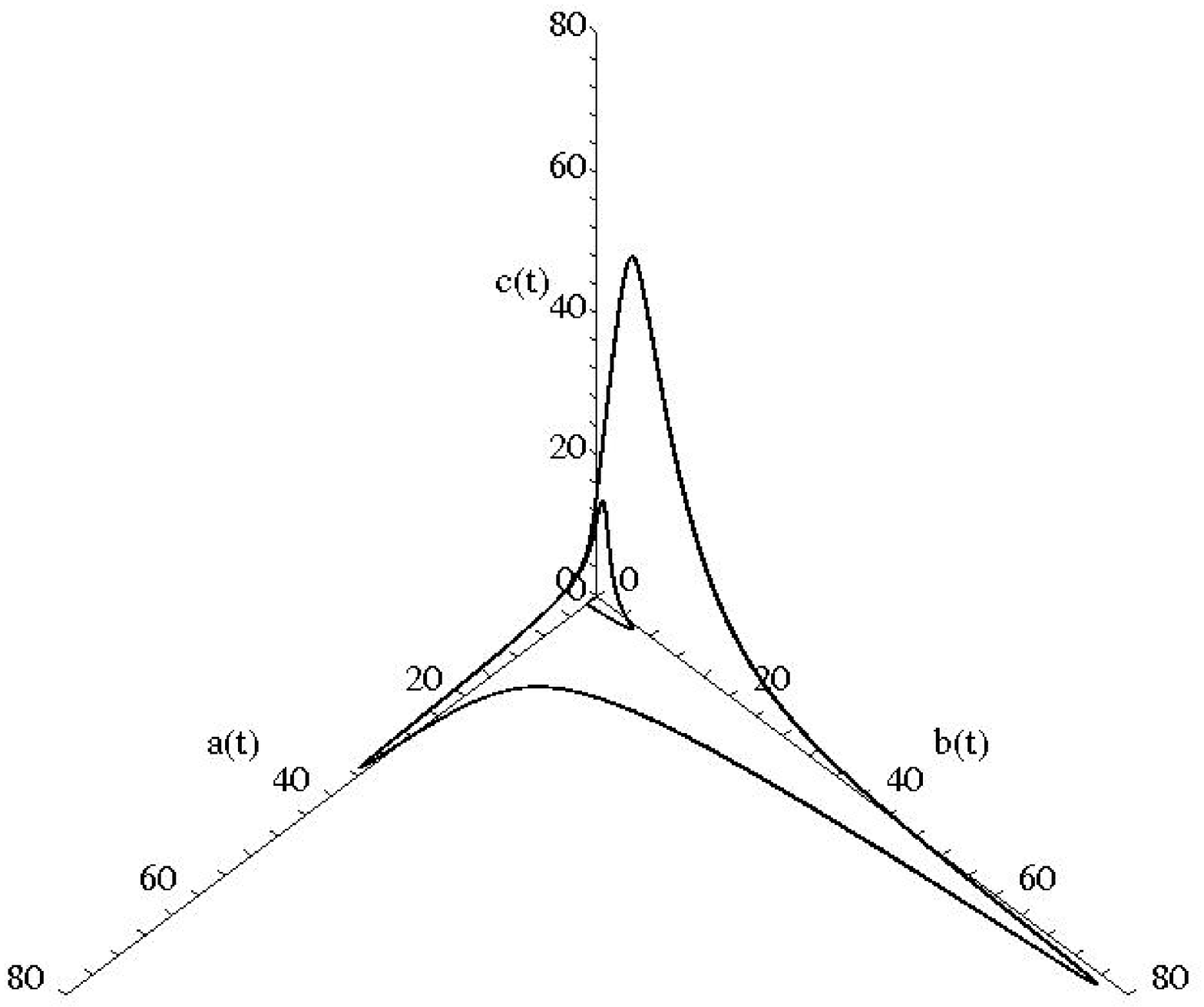}
\end{center}
\vspace{-2cm}
\caption{(Color online) 
Top: the rewired repressilator: a positive loop is added (see arrow), so that one of the
genes is doubly regulated. Bottom: the limit cycle of the (reduced) rewired repressilator circuit; the additional activation interaction 
breaks the symmetry, as discernable in the difference in maximal concentrations. Simulation parameters are: $r $ =1, $r_p $ =  $ 10^{-4} $,  $\varepsilon$ = 0.1, $\eta_1=  \eta_2$=  $ 10^{-2}$, $\gamma = 10^{-3}$, $h = 3$).}
\end{figure}
In the case without genes, this means that one of the equations, say the one for $a$
is replaced by
\begin{equation}
\dot{a} = \frac{\varepsilon + r_p c^h}{1 + \nu b^h + \nu_p c^h} - \gamma a
\end{equation}
This `rewired' repressilator still has a unique fixed-point $(a,b,c)$, as follows from an analysis of the
fixed-point conditions. The stability condition can be read off, as before, from the stability
matrix which now reads as
\begin{equation}
\Gamma^{fp}
= 
\left(
\begin{array}{ccc}
-\gamma   & - \kappa_0               &  \kappa_1       \\
0                 & -\gamma                  &  - \kappa_2         \\
- \kappa_3     &  0                              & - \gamma  
\end{array}
\right)
\end{equation}
with 
\begin{equation}
\kappa_0 \equiv 
\frac{\nu h b^{h-1}(\varepsilon + r_p c^h)}{(1 + \nu b^h + \nu_p c^h)^2}\, ,
\end{equation}
\begin{equation}
\kappa_1 \equiv
\frac{h c^{h-1}(r_p(1 + \nu b^h) - \nu_p\varepsilon)}{(1 + \nu b^h + \nu_p c^h)^2} \, , 
\end{equation}
\begin{equation}
\kappa_2 \equiv - \frac{\nu h c^{h-1}}{(1 + \nu c^h)^2} \, ,
\end{equation}
\begin{equation}
\kappa_3 \equiv - \frac{\nu h b^{h-1}}{(1 + \nu b^h)^2} \, .
\end{equation}
Note that $\kappa_1$ can be both positive and negative. 
The characteristic polynomial reads 
\begin{equation}
(\gamma + \lambda)^3 + (\gamma + \lambda) \kappa_1\kappa_3  + \kappa_0\kappa_2\kappa_3 = 0   
\end{equation}
which still has a pair of complex eigenvalues. The Hopf condition is given by
\begin{equation}
8 \gamma^3 + 2\gamma\kappa_1\kappa_3 - \kappa_0\kappa_2\kappa_3 = 0\, .
\end{equation}

The analysis of the full system, genes included, is clearly more involved than for the repressilator 
due to the increased number of variables. We have therefore studied the system only numerically and 
compared the reduced and the full version, as we did for the repressilator. 
Our calcuations show that the reduced version (3 ode's for $a$,$b$,$c$) 
is less robust against rewiring than the gene gate version (7 ode's): the stability limit of 
the limit cycle regime can differ by parameter values up to one order of magnitude. 
This finding is notable since in the presence of multiple regulations the number of gene
states increases linearly with the number of inputs (neglecting still additional regulatory
layers) and thus significantly enhances the complexity in modeling circuits with such
elements. We close the section with Figure 7 (bottom) which shows the limit cyle of the rewired repressilator for the reduced deterministic system ($ h = 3 $). 
It illustrates that in general the presence of the additional positive loop breaks the ($a-b-c$)
symmetry between concentrations. 

\subsection{A multi-input circuit related to developmental regulation}

In this final subsection we address a second example of a multi-input gate. It consists of a bistable switch
built from two repressing gates which is placed under additional control by an activating input. Such 
motifs occur both in transcriptional regulation \cite{babu03}, but they have also been
proposed recently to play a role in morphogen concentration-dependent cellular development
\cite{saka07}; our example is motivated by the latter case. 
The circuit dynamics is governed by the following ode's (neglecting
the gene gate dynamics since we are concerned with fixed-point dynamics only)
\begin{eqnarray}
\dot{b} & = & \frac{\varepsilon_b + r a^n}{1 + \nu c^m + \nu_{ac} a^n} - \gamma b \\
\dot{c} & = & \frac{\varepsilon_c + r a^n}{1 + \nu b^l + \nu_{ab} a^n}  - \gamma c
\end{eqnarray}
where $m,n,l$ are the different Hill exponents. If the activating variable $a = 0$, the system
is the standard bistable switch, albeit asymmetric with respect to the parameters and nonlinearities,
and it is this asymmetry which plays an important role - in ref. \cite{saka07}, the supposed Hill coefficients have values of 3 and 6, respectively. 

The effect of the variable $a$ has on the dynamics is easily understood. To simplify 
matters, we neglect $a$ in the first equation and look at an asymmetric wiring. 
It actually does not matter whether we allow $a$ to control one or both transcription factors
$b,c$ as long as $a$ interacts with both in the same way and not via a different nonlinearity: the
main symmetry-breaking effect is contained in the difference between the Hill coefficients 
controlling $b$ and $c$. 

Supposing further that we increase the concentration of $a$ 
to levels where it dominates the concentration $b$ so that we have for the fixed-point in $c$
\begin{equation}
c_0 = \frac{1}{\gamma}\frac{\varepsilon_c + r a^n}{1 + \nu_{ab} a^n} 
\rightarrow_{a \gg 1} \frac{\eta_{ab}}{\gamma}
\end{equation}
Thus, the fixed-point concentration of the repressing variable $c_0$ is locked to that of $a$ and 
approaches an asymptotically constant value. Correspondingly, this brings the fixed-point level of
$b$ down and under firm control of $a$: the system ceases to be bistable, and locks
into a stable state under control of $	a$. The possible relevance of this mechanism 
for a transcriptional circuit in development is evident: increasing $a$ can
force the system to switch in a concentration-dependent way. 

In the nonlinear dynamics case, this switch is therefore brought about by the vanishing of a fixed-point;
again, this situation is different in the stochastic setting.
For comparison, Figures 8 and 9 show our results of the stochastic simulations for the  circuit 
\begin{equation}
null(a)|posneg(a,b;c)|posneg(a,c;b) 
\end{equation}
without any cooperative nonlinearity, as for the repressilator. 
\begin{figure}
\begin{center}
\includegraphics[width=0.4\textwidth]{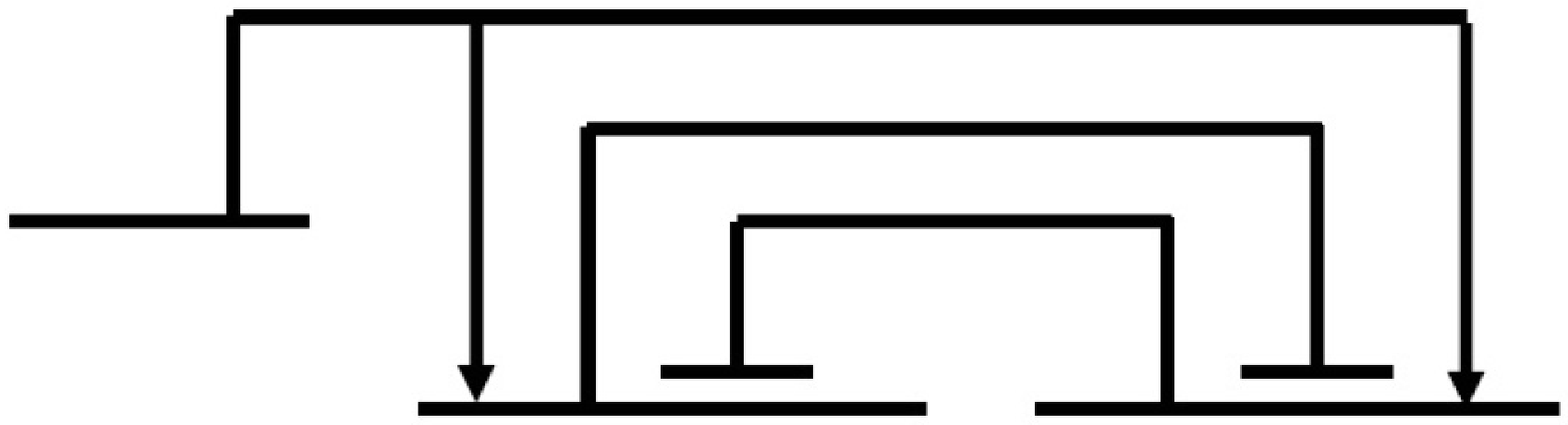}
\includegraphics[width=0.5\textwidth]{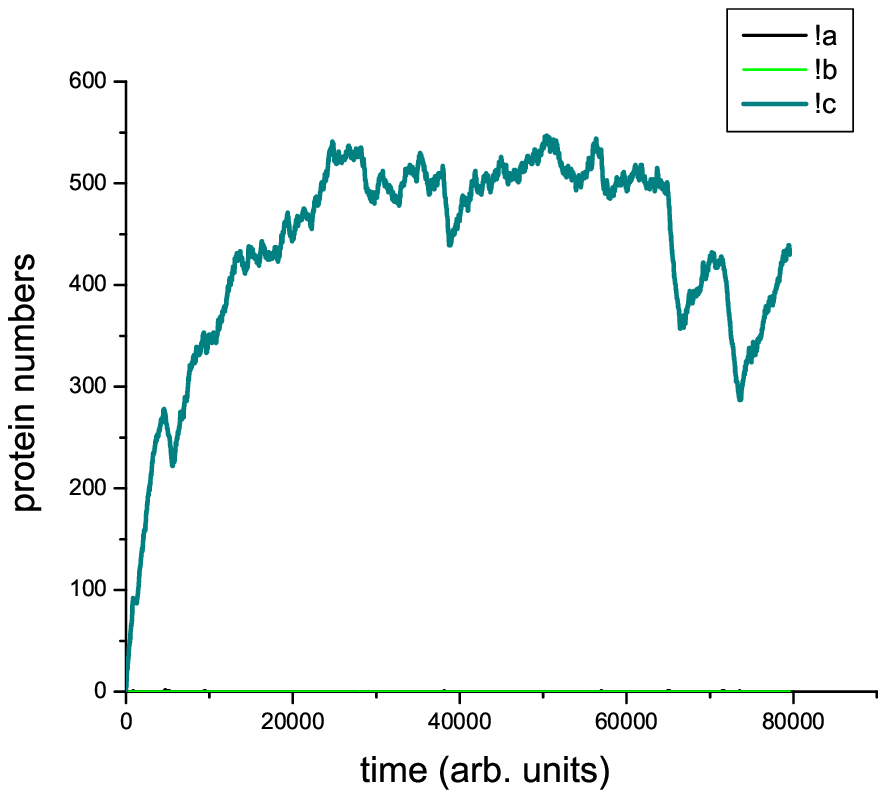}
\end{center}
\caption{(Color online) 
Top: the bistable switch under external control by $a$. Bottom: the system starts at zero concentrations of both proteins and enters the state with higher stability, as given by higher production rates. The signal $c$ dominates widely over $a,b$ which are indistinguishable from the baseline.}
\end{figure}
\begin{figure}
\begin{center}
\includegraphics[width=0.5\textwidth]{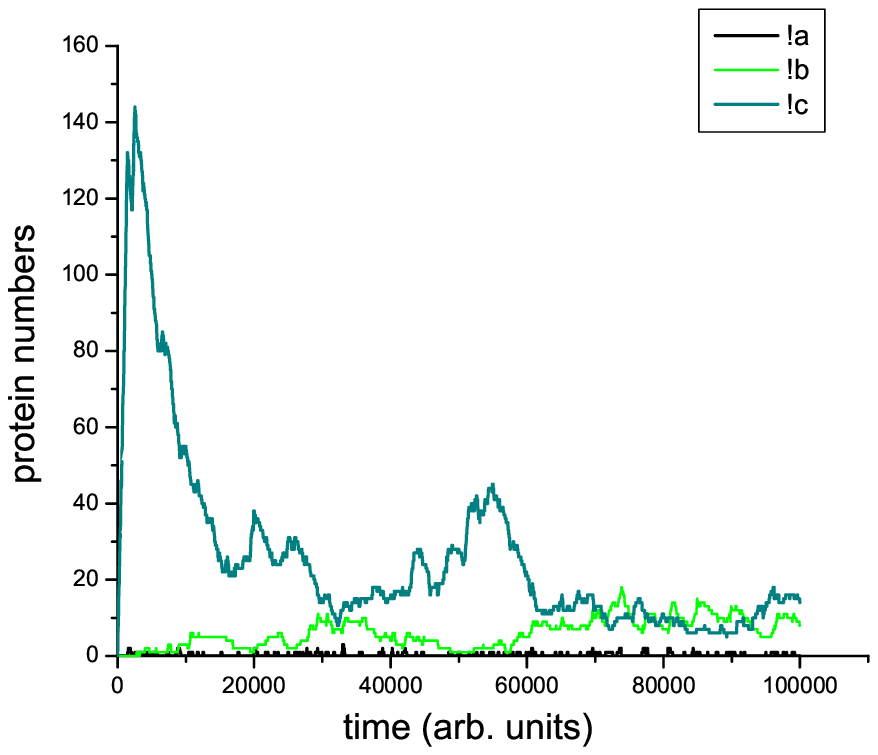}
\includegraphics[width=0.5\textwidth]{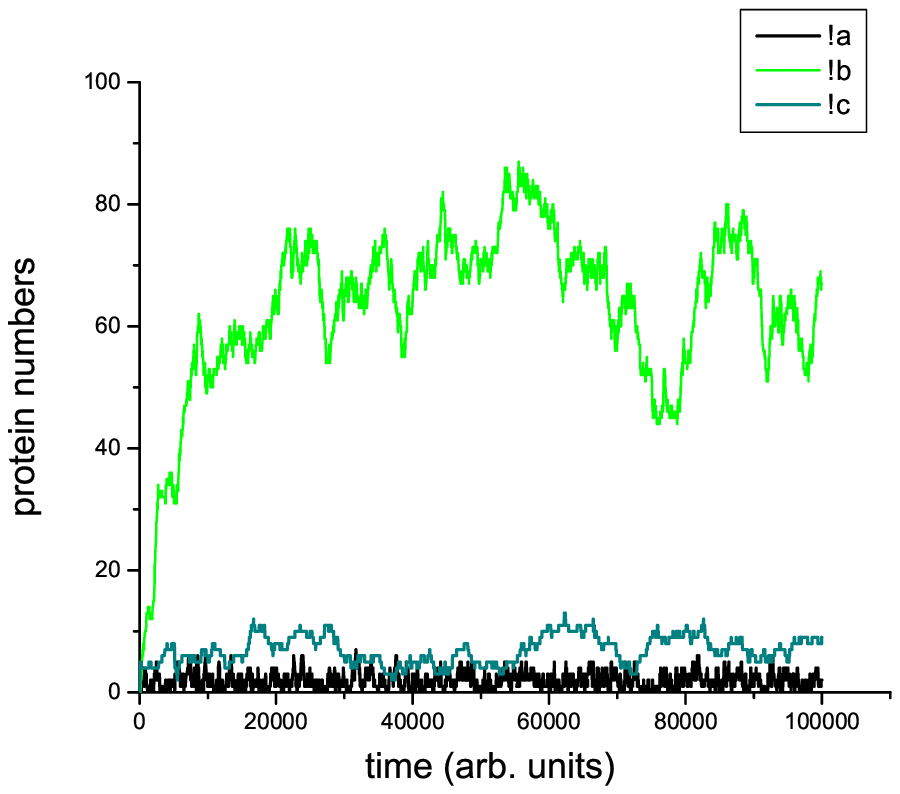}
\end{center}
\caption{(Color online)
Top: a change of the activator transcription rate by one order of magnitude makes both
proteins compete; note the concentration overshoots of the previously stable protein. Bottom:
A further increase of the activator transcription rate makes the system switch between the two states.
Simulation parameters are (bottom): $r$ =1,  $\varepsilon_a$ = $10^{-3}$, $\varepsilon_b$ = 0.1,            $\varepsilon_c$ = $10^{-3}$, $\eta_{ab}$= 2$\cdot 10^{-3}$,  $\eta_{ac}$ = 2$\cdot 10^{-2}$, 
$ \eta$ = 2$\cdot 10^{-1}$, $\gamma_a$ = 5$\cdot 10^{-3}$, $\gamma$ = 2 $\cdot 10^{-4}$.}
\end{figure}
The progression of dynamic behaviours in Figure 8 bottom to Figure 9 top and bottom 
is controlled by the average concentration level
of $a$, which increases from one figure to the next by one order of magnitude since the 
transcription rate is increased by this factor. In Figure 8 (bottom) 
the switch enters the more stable of the two states; in Figure 9 (top) the additional input $a$ makes 
the concentration levels $b$ and $c$ compete with each other. This behaviour is observed 
within a large parameter range, in which bursts in concentration $c$ can occur at random times 
within a wide time interval (see the concentration peak at around 55.000 a.u.), 
and are finally controlled by $a$. In Figure 9 (bottom) the system has switched to
a dominant concentration of $b$ and the concentration of the previously dominant transcription 
factor $c$ is now fully controlled by $a$.  Note the difference in concentration levels of all proteins 
in the figures.

\section{Discussion and Outlook}

In conclusion we have proposed a minimal model description for gene regulatory networks
based on the notion of the gene gate, first proposed in ref. \cite{blossey06}. 
We studied the dynamics of simple gene networks in both a mean-field and a stochastic
version, with characteristically different results: 
\begin{itemize}
\item If the system dynamics is stable fixed-point only, a reduced deterministic description 
ignoring the degree of freedom of the gates is sufficient in the sense that the fixed-point
is not altered by the presence of the genes. But if this is the case, the latter are indeed
`irrelevant'. In order to represent faithfully the fixed-point structure of the network, a Hill-type 
nonlinearity may be needed (like for the bistable switch circuit). 
However, within a stochastic description fluctuation effects induced by the genes 
(promoters) might affect fixed-point locations \cite{paulsson00}, or the stability of bistable switches 
\cite{warren04,allen05,walczak05,morelli08}.
\item If the system displays a limit cyle, the gene gates are relevant, as is any other
additional regulatory layer to determine the parameter range of oscillations. In general
the limit cycle regimes depends on the whole set of parameter values, Hill coefficient 
included. In particular this means that in multi-input regulations in which additional 
gene states have to be accounted for, the parameter space can extend significantly. 
\item If the system dynamics is fixed-point, the stochastic version obeys this without any
need for cooperative effects. The same holds true for limit cyle behaviour. Additional
regulatory layers also enlarge the phase space but in a trivial way. By contrast, they
affect oscillatory behaviour by regularizing the oscillations. 
\end{itemize}
In our view these results have interesting consequences on the philosphy of modeling
gene regulatory networks in suggesting a different coarse-approach. Computational
models of large networks can be built by abstracting away all regulatory layers to a
level where the remaining network can still faithfully represent the system characteristics.
Network motifs that have a more sensitive dynamic behaviour - like limit cycles, as shown
here - are more sensitive to modeling assumptions. Finally, we remark that in view of
our results, modeling attempts combining deterministic and stochastic aspects should
be considered with care \cite{scott07}.
\\
\begin{acknowledgements}
We thank Luca Cardelli, Andrew Phillips and Yasushi Saka for discussions.
\end{acknowledgements}

\section{Appendix: Simulations in stochastic $\pi$-calculus}

The Gillespie algorithm can, of course, be implemented in various different programming
languages. What then are the main ideas and advantages of the $\pi$-calculus? 
 
The $\pi$-calculus is a formal system in which each computation is represented by a
communication over input and output channels. The communicating objects are
called `processes'. Computation by communication within pi-calculus can be understood
as an alternative to, e.g., functional computation as realized in the $\lambda$-calculus.
The $\pi$-calculus is Turing complete: it can therefore realize any possible computation
\cite{milner99}.

For our application, the calculus allows to represent each gene gate by a computational process
\begin{equation}
gate(x;y) 
\end{equation}
with its corresponding input(s) $x$ and output(s) $y$; e.g. the repressing gate of Figure 1.2 is written as
$ neg(a;b) $ where the input channel $a$ represents the repression of transcription by transcription 
factor $a$, and $b$ is the corresponding output. All other reactions, like e.g. the degradation 
process of $b$, are bound to this process and contained in its definition. 

The scheduling of inputs and outputs on a gate are calculated in the usual fashion by the 
standard Gillespie algorithm, as adapted to the  $\pi$-calculus 
\cite{phillips07,blossey06,blossey07}. 

One main technical advantage of the calculus is, in fact, that its syntax and semantics are perfectly adapted to a `compositional' build-up of the transcriptional networks. In our
context this permits to express (and compute!) a composed circuit, like the repressilator, 
by a parallel process 
\begin{equation}
neg(c;b)|neg(b;a)|neg(a;c)
\end{equation}

The second main advantage (although not exploited for the small systems studied here) is
that it can reduce the computational complexity of a system of $n$ kinetic reactions, which
is of order $n^2$, to linear order. 
The interested reader is referred to refs. \cite{blossey06,blossey07} for more details, written
in a way accessible to a physics-trained audience.
The simulation results presented here were obtained with the public domain software SPIM, downloadable with 
documentation and examples \cite{phillips08}.  The details of the implementation of the Gillespie 
algorithm in the dedicated software SPIM are discussed in the Supplementary Materials of  
\cite{blossey06,blossey07}.

\end{document}